\documentclass[fleqn,10pt]{wlscirep}
\usepackage[pdftex]{dropping}

\title{Tracking of plus-ends reveals microtubule functional 
diversity in different cell types}

\author[1,*,$\dagger$]{M. Reza Shaebani}
\author[2,*]{Aravind Pasula}
\author[2]{Albrecht Ott}
\author[1]{Ludger Santen}
\affil[1]{Department of Theoretical Physics, Saarland University, 66041 
Saarbr\"ucken, Germany}
\affil[2]{Department of Experimental Physics, Saarland University, 66041 
Saarbr\"ucken, Germany}
\affil[*]{These authors contributed equally to this work.}
\affil[$\dagger$]{Corresponding author. shaebani@lusi.uni-sb.de}

\begin{abstract}
Many cellular processes are tightly connected to the dynamics 
of microtubules (MTs). While in neuronal axons MTs mainly 
regulate intracellular trafficking, they participate in 
cytoskeleton reorganization in many other eukaryotic cells, 
enabling the cell to efficiently adapt to changes in the 
environment. We show that the functional differences of MTs 
in different cell types and regions is reflected in 
the dynamic properties of MT tips. Using plus-end tracking 
proteins EB1 to monitor growing MT plus-ends, we show that 
MT dynamics and life cycle in axons of human neurons 
significantly differ from that of fibroblast cells. The 
density of plus-ends, as well as the rescue and catastrophe 
frequencies increase while the growth rate decreases toward 
the fibroblast cell margin. This results in a rather stable 
filamentous network structure and maintains the connection 
between nucleus and membrane. In contrast, plus-ends are 
uniformly distributed along the axons and exhibit diverse 
polymerization run times and spatially homogeneous rescue 
and catastrophe frequencies, leading to MT segments of 
various lengths. The probability distributions of the 
excursion length of polymerization and the MT length both 
follow nearly exponential tails, in agreement with the 
analytical predictions of a two-state model of MT dynamics.
\end{abstract}

\begin{document}

\flushbottom

\maketitle

\thispagestyle{empty}

\dropping[0pt]{2}{M}icrotubules are semiflexible polymers with 
an intrinsic structural polarity. They represent tracks for the 
transport of material within the cell by means of molecular motor 
proteins. Active transport is essential for an efficient delivery 
of cargoes to specific locations through the crowded cytoplasm 
\cite{Bressloff13}, and several types of diseases arise due to 
perturbations in intracellular transport processes. The dynamic 
structure of microtubules (MTs) has been suggested to be beneficial 
for reducing jam formation and maintaining homogeneous states in 
bidirectional transport of molecular motors \cite{Ebbinghaus11}. 
The transport efficiency may be dramatically affected by the 
drugs which stabilize (e.g.\ taxanes) or destabilize (e.g.\ 
vinca alkaloids) MT structure \cite{Stanton11}. Besides the role 
of MTs in material delivery, their dynamics enables the cells to 
quickly remodel their cytoskeleton in response to environmental 
changes \cite{Pullarkat07}. This leads to an efficient control 
of vital processes such as mitosis and cell division, motility, 
and morphogenesis. In cell types that benefit from the presence 
of MTs to adjust their morphological requirements, having a stable 
MT network near the cell margin is necessary, in contrast to cell 
types where MTs are not involved in the steady remodeling of the 
cell shape. In such cases, for example in neuronal axons, a more dynamic 
MT structure may be even more advantageous because of enhancing 
the transport capacity. The ability of MTs to rapidly switch 
between growth and shrinkage states, known as \emph{dynamic 
instability} \cite{Mitchison84}, is assigned to the complex 
interplay between the applied stresses \cite{Dumont09,Goshima10}, 
GTP hydrolysis \cite{Mitchison84}, and regulatory proteins 
including molecular motors \cite{Vorobjev01,Melbinger12,
Johann12,Kinoshita01} and MT-associated proteins 
\cite{Kinoshita01,Howard07,Wordeman05,Tournebize00}. 

In the present study, we demonstrate that the differences in 
MTs function in different cell types and regions is reflected 
in their dynamic structure. To this aim we label MT plus-ends 
by means of EB1 proteins and compare their motion in fibroblast 
and human neuronal cells. Transfection of human neurons is 
technically challenging, however, uncovering the details 
of MT dynamics in such cells is of crucial importance to 
diagnose, treat, or even prevent neurodegenerative disorders. 
MT dynamics have been studied in different nerve cells such 
as mouse hippocampal \cite{Stepanova03}, Aplysia 
\cite{Shemesh08} and Drosophila \cite{Stone08} neurons. 
We report, for the first time, the structural properties 
of MTs in human neuronal cells by culturing SH-SY5Y cells, 
a well documented human derived neuroblastoma cell line which 
differentiates to mature neurons after treatment with All-trans 
retinoic acid and brain-derived neurotrophic factor 
\cite{Encinas00}. The SH-SY5Y cell line is widely used as 
an in vitro model to study biochemical and functional 
properties of neurons. We clarify the differences between 
axonal MT polymerization/depolymerization excursions as 
well as the spatial homogeneity of their plus-end tips 
with those of fibroblast cells. 

The MT-associated proteins (MAPs) may stabilize or destabilize MTs 
in living cells by temporally or spatially regulating their dynamics. 
MAPs target MT-ends and/or walls, or the non-polymerized tubulin 
subunits. Among various types of MAPs, the plus-end tracking proteins 
(+TIPs) accumulate at growing MT plus-ends and play important roles 
e.g.\ in regulation of MT dynamics, delivery of signaling molecules, 
and control of MT interactions with other intracellular structures 
\cite{Akhmanova08,Wu06,Akhmanova05,Maurer12}. +TIPs may interact 
with each other and construct plus-end complexes. Particularly, 
the end-binding protein-1 (EB1) is frequently involved in such 
complex structures \cite{Dixit09}. EB1 is a member of dynamic and 
enigmatic family of +TIPs, which is highly conserved from humans 
to yeasts and plants, and acts as an exquisite marker of dynamic 
MT plus-ends \cite{Tirnauer00,Schuyler01}. EB1 senses conformational 
changes, which occur in the MT lattice, linked to the GTPase cycle 
of tubulin at growing MT ends \cite{Maurer11}. This leads to the 
autonomous comet-like accumulation of EB1 at the growing MTs. 

In axons MTs are generally oriented, with their plus (minus) ends 
pointing toward the axon terminals (the soma) \cite{Stone08,Conde09}. 
In contrast to many eukaryotic cells in which the minus ends of MTs 
are mainly anchored at the MT organizing center, MTs do not reach 
from soma all the way to axon terminals in neurons. Instead, there 
is an overlapping array of short segments of MT with a typical 
length scale of a few micrometers. 

Here, we extract the length distribution $P(L)$ of MT segments from 
the spatial distribution of the labeled plus-ends and show that the 
tail of $P(L)$ decreases nearly exponentially. By means of a 
two-state model of MT growth and shrinkage, it is demonstrated how 
the steady-state length distribution depends on the phenomenological 
parameters: the growth and shrinkage rates and the frequency of 
catastrophe and rescue events, i.e.\ switching between growth and 
shrinkage states and vice versa. 

\section*{Methods}
\label{Sec:Methods}
\subsubsection*{Cell culture and differentiation} 
SH-SY5Y cells were cultured in growth medium containing Dulbecco's 
Modified Eagle Medium (DMEM; Gibco) supplemented with $10\%$ 
heat-inactivated Fetal Calf Serum (FCS; PAA Laboratories, Austria), 
50 U/ml penicillin 50$\,\mu$g/ml streptomycin (Sigma Aldrich) 
and 2 mM L-glutamine (Sigma Aldrich). NIH swiss 3T3 cells (DSMZ, 
Germany) were grown in DMEM supplemented with $10\%$ 
FBS, 2 mM L-glutamine, and $100\,\text{U}/\text{ml}$ penicillin 
100$\,\mu$g/ml streptomycin. Both the cell types were cultivated 
in T25 flasks at $37^{\circ}$C, humidified air with $5\%$ $\text{CO}_2$. 
The medium was changed regularly twice a week and the cells were split 
before they reached confluence.

For microscopy, the SH-SY5Y cells were cultured in $35 \text{mm}$ 
$\mu$ dishes (ibidi) which were previously coated with 
$50\,\mu$g/ml collagen (Corning). Cells were differentiated the day 
after plating by $10\,\mu\text{M}$ All-trans retionoic acid (RA; 
Sigma Aldrich). After 5 days, the cells were washed three times with 
DMEM and grown in serum-free DMEM supplemented with $50\,$ng/ml 
brain-derived neurotrophic factor (BDNF; Sigma Aldrich) for 7 days. 
After treatment, cells exhibit biochemical and morphological 
features similar to those of mature human neurons \cite{Encinas00}. 

NIH-3T3 cells were cultured in $35 \text{mm}$ imaging dishes pre-coated 
with 10$\,\mu$g/ml fibronectin (Sigma Aldrich) for microscopy. Both 
the cell lines were liposome-transfected by pGFP-EB1 plasmid \cite{Piehl03} 
(Addgene 17234) by means of the torpedo DNA transfection vector (Ibidi) 
in DMEM media without sera and antibiotics according to manufacturer's 
protocol. Live cell microscopy started 18-48 hrs after transfection.

\subsubsection*{Live cell imaging and processing} 
Cells successfully expressing GFP were chosen and analyzed with an 
Axio observer Z1 inverted fluorescence microscope equipped with an 
Axiocam Mrm, Incuabator X1multi S1, TempModul S1, and $\text{CO}_2$ 
Modul S1 (all from Zeiss). Images were taken every 1s at an exposure 
time of 600-800 ms with a 488nm laser for 15-20 min with $100\times$ 
objective. All the measurements were performed in a humidified atmosphere 
at $37^{\circ}$C and at $5\%$ $\text{CO}_2$. Quantitative analysis of 
the microtubule dynamics was carried out on time-lapse movies of cells 
expressing EB1-GFP. Microtubule growth rates were obtained by tracking 
EB1-GFP comets at microtubule plus-ends. Images were recorded and 
movies were assembled by means of AxioVision software. More than 750 
MT tips in Fibroblast cells and nearly 800 MT tips in axons of human 
neurons were analyzed.

\begin{figure}
\centering
\includegraphics[width=0.95\textwidth]{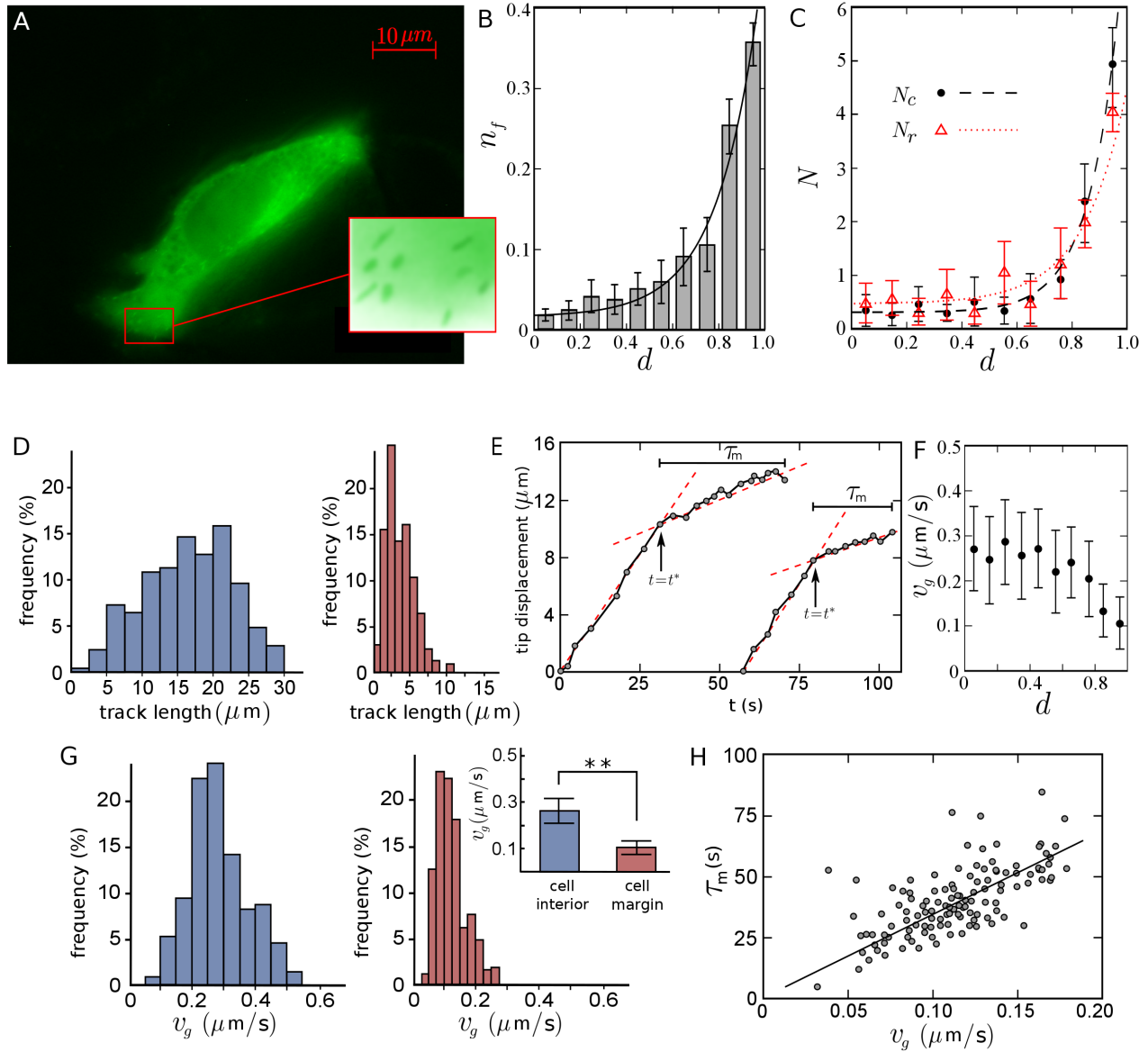}
\caption{MT dynamics in fibroblast cells. (A) Distribution of EB1 
labeled microtubule tips (spots) in a fibroblast cell. The inset 
shows a zoomed part of the cell margin. (B) The fraction $n_f$ of 
active plus ends versus $d$, the relative position of MT tip between 
the cell interior and margin. The solid line indicates the best 
fit to an exponential growth $n(d){\sim}\exp[k\, d]$ with $k{=}
4.36\pm0.37$. The imaging frequency is $1\,\text{frame}{/}
\text{s}$ and the total observation time is $850$ s. (C) The 
number of rescue $N_r$ and catastrophe $N_c$ events (per 100 
growing tips) versus $d$. The dashed and dotted lines denote 
exponential fits with constants $9.06{\pm}0.66$ and 
$6.81{\pm}1.61$, respectively. (D) Frequency histogram 
of the excursion length of polymerization (i.e.\ the run length 
in the growth phase) in the cell interior ($d{<}0.8$, left) and 
near the cell margin ($d{>}0.8$, right). (E) Temporal evolution 
of the position of a few typical plus-end tips that reach the 
cell margin. The dashed lines are obtained from the three-parameter 
fit, as explained in the text. (F) The growth velocity $\text{v}_g$ 
versus $d$. (G) Frequency histogram of the growth velocity in 
the cell interior (left) and near the cell margin (right). The 
inset shows the corresponding mean values ($P{\leq}0.01$, t 
test). (H) Scattered data points showing the excursion time of 
polymerization near the cell margin, $\tau_m$, versus the growth 
velocity, $\text{v}_g$. The solid line indicates the best linear fit.}
\label{Fig1}
\end{figure}

\section*{Results}
\label{Sec:Results}

\subsubsection*{Characterization of MT dynamics in fibroblast cells}
\label{Subsec:FibroblastCells}
We first study fibroblast cells, in which the dynamic behavior of 
MTs is essential for cytoskeletal reorganization.  We measure the 
phenomenological characteristics of MT dynamics such as the growth 
velocity and the frequencies of rescue or catastrophe events, and 
clarify their differences in the cell interior compared to the 
cell margin. In the model section we demonstrate how these 
differences correspond to different MT growth strategies and lead 
to distinct steady-state MT lengths. 

The distance of MT tip from the cell margin is a decisive 
parameter in determining its dynamics. To investigate how MT 
dynamics changes when approaching the fibroblast cell margin, we 
characterize the location of each MT plus-end with respect to the 
centrosome and plasma membrane by a dimensionless quantity $d$ 
ranging from $0$ (center) to $1$ (margin). In case of elongated 
fibroblast cells, $d$ denotes the relative distance of the MT tip 
to the plasma membrane parallel to the direction of elongation of 
the cell. Otherwise, $d$ is calculated for each individual MT 
tip based on its current position (i.e.\ in the chosen image frame) 
along its trajectory. We checked that other choices to characterize 
the location of the MT tip with respect to the cell margin, such 
as the absolute distance from the membrane, lead to qualitatively 
similar results for the behavior of MT dynamics parameters as 
a function of the location of the MT tip. The analysis of live-cell 
images reveals that the plus ends are more concentrated in the 
vicinity of the cell margin (see Figs.~\ref{Fig1}A and \ref{Fig1}B). 
The increase of the fraction of active tips $n_f$ with the relative 
distance $d$ is accelerated towards the plasma membrane, which 
can be quantitatively described by an exponential growth in $d$. 
The question arises how the cell manages to adjust the spatial 
distribution of the MT tips. To address this, we measure the 
accessible quantities related to the dynamics of MTs and 
consider their evolution as a function of the distance from 
the membrane, which provides a better understanding of the 
underlying mechanisms of MT length regulation. 

The relatively high density of tips near the membrane is a signature 
of unstable dynamics and more frequent switching between the growth 
and shrinkage phases, compared to the cell interior. We directly 
examine this by measuring the number of catastrophe or rescue
transitions, i.e.\ switching events from growth to shrinkage phase 
and vice versa, respectively. In order to reduce the image analysis 
errors, a minimum life-time threshold of two successive frames 
after creation (before disappearance) is imposed on a signal to 
consider the event as a rescue (catastrophe) transition. The 
number of rescue $N_r$ and catastrophe $N_c$ events per each 
100 growing tips are shown in Fig.~\ref{Fig1}C. Both $N_r$ 
and $N_c$ increase towards the cell membrane which can be described 
by exponential functions in terms of $d$, where the increase of 
catastrophe events is more pronounced. Thus, with approaching the 
membrane, MTs more frequently experience switching events, which 
prevents long excursions of polymerization/depolymerization and leads 
to a high concentration of active tips in the vicinity of the plasma 
membrane. Figure \ref{Fig1}D shows that the probability distribution 
of the excursion length of polymerization $\ell_g$ is broader 
in the cell interior and its average value is larger. When separating 
the cell interior and margin with the threshold value $d{=}0.8$ 
(see below), we obtain $\langle \ell_g \rangle {\simeq} 17.8\,\mu\text{m}$ 
and $4.2\,\mu\text{m}$, respectively, for the cell interior and margin.

We also identify individual MTs and follow their trajectories 
to see how the tip proceeds when it approaches the membrane. In 
Fig.~\ref{Fig1}E, tip displacements of a few typical MTs are shown. 
It can be seen that the growth velocity remains nearly unchanged 
in the cell interior (provided that it does not face large 
obstacles which are spatially constrained), but it is drastically 
reduced near the plasma membrane. By measuring the instantaneous 
growth velocity $\text{v}_g$ as a function of the dimensionless 
quantity $d$, it is shown in Fig.~\ref{Fig1}F that $\text{v}_g$ 
is considerably lower near the membrane. The sharp change of the 
growth velocity for individual MTs provides the opportunity to 
quantitatively discriminate between the cell interior and margin. 
To this aim, we fit the time evolution of the position of the tip 
with the following function
\begin{eqnarray}
d(t)\!\!=\!\! \left\{ \begin{array}{ll}
     \text{v}\!_{_g}^{\;i} \, \cdot t & \mbox{\;\;\;\;\;\;\;\;\;$t < t^*$} \\
     \text{v}\!_{_g}^{\;m} \! \cdot t & \mbox{\;\;\;\;\;\;\;\;\;$t > t^*$}
     \end{array},
     \right.
\label{Eq:ThreeParameterFit}
\end{eqnarray}
where the three free parameters $\text{v}\!_{_g}^{\;i}$, 
$\text{v}\!_{_g}^{\;m}$, and $t^*$ respectively denote the 
mean growth velocity in cell interior, mean growth velocity 
in cell margin, and the onset of transition from interior to 
margin. By varying $t^*$ we minimize the fitting errors and 
obtain the best set of fit parameters (see Fig.~\ref{Fig1}E). 
Repeating this procedure for more than 300 MT tips, the frequency 
histogram of the growth velocity is separately obtained for 
the cell interior and margin (see Fig.~\ref{Fig1}G). The 
corresponding average values are $\text{v}\!_{_g}^{\;i}{=}
0.28{\pm}0.05 \,\mu\text{m}{/}\text{s}$ and $\text{v}\!_{_g}^{\;m}
{=}0.09{\pm}0.03 \,\mu\text{m}{/}\text{s}$. Moreover, the transition 
point is, on average, located at $d{\approx}0.77$. We also measure 
the excursion time $\tau_m$ of polymerization near the cell margin. 
The resulting values are plotted in terms of $\text{v}\!_{_g}^{\;m}$ 
in Fig.~\ref{Fig1}H. Interestingly, we observe a positive correlation 
between these quantities (their Pearson correlation coefficient is 
nearly $0.57$); the faster the MT tip is, the longer it survives 
in the growing phase.    

\subsubsection*{MT dynamics in axons of human neurons}
\label{Subsec:Axons}
\begin{figure}[t]
\centering
\includegraphics[width=0.8\textwidth]{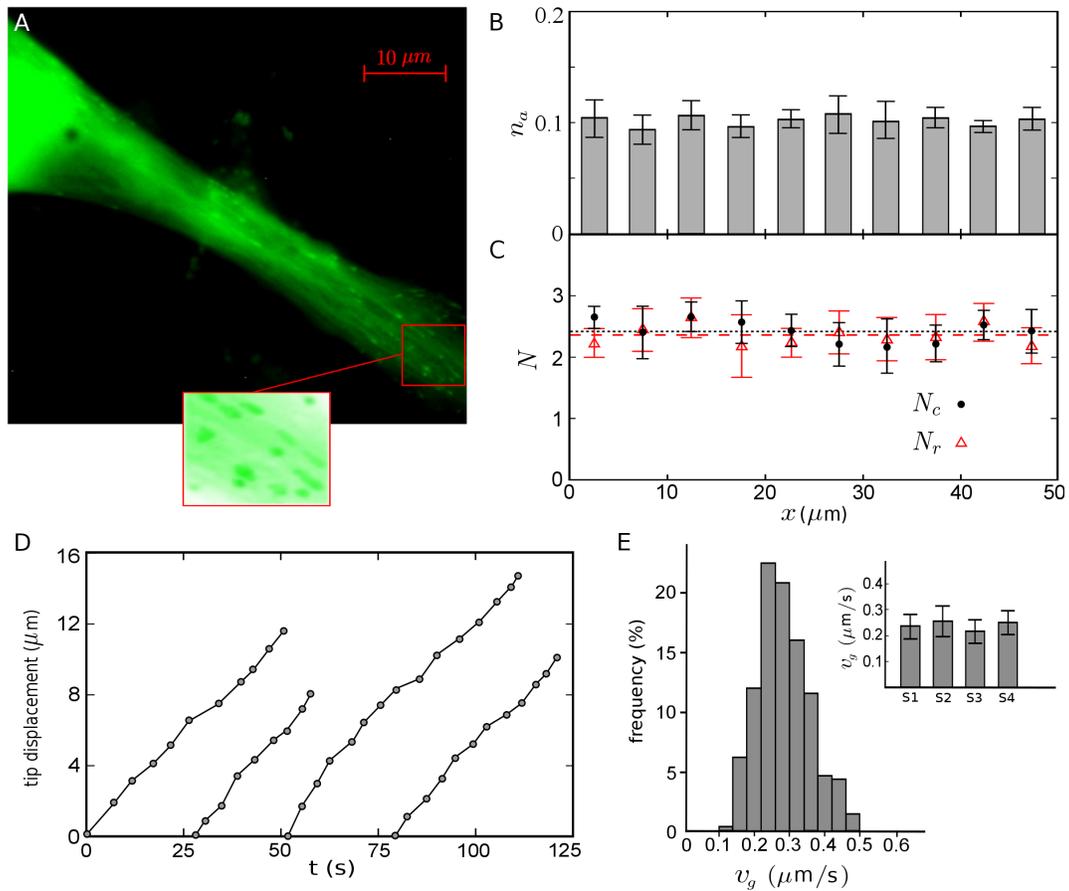}
\caption{MT dynamics in axons of human neurons. (A) Distribution 
of EB1 labeled microtubule tips (spots) in axons. The inset shows 
a zoomed part of the axon. (B) The linear density $n_a$ of active 
plus ends (corrected for the variations of the cross-section 
area of axon) versus the distance $x$ from the soma. A distance 
of $50\,\mu\text{m}$ is considered. The imaging frequency is 
$1\,\text{frame}{/}\text{s}$ and the total observation time is 
$850$ s. (C) The number of rescue $N_r$ and catastrophe $N_c$ 
events (per 100 growing tips) versus the distance $x$ from the 
soma. The dashed lines indicate mean values. (D) Temporal 
evolution of the position of a few plus-end tips along the axon. 
(E) Frequency histogram of the growth velocity $\text{v}_g$. 
Inset: The mean value of $\text{v}_g$ obtained from four different 
axons.}
\label{Fig2}
\end{figure}

Next, we investigate MT dynamics in axons of neurons. In the 
absence of MT organizing center, segments of MTs are distributed 
nearly parallel to the plasma membrane. As MT tips do not 
contribute in keeping the shape of axons, it is expected that 
the parameters describing the dynamic structure of MTs are 
rather homogeneously distributed. Taking the variations of 
the thickness of axon tube into account, we find that the 
linear density of active tips is uniform along the axon (see 
Fig.~\ref{Fig2}B), evidencing that the distance from the 
soma is not an influential parameter for MT dynamics. Our 
detailed analysis also showed no significant difference in 
the density of plus-end tips across the cross-section of 
axon (not shown). Additionally, it can be seen from 
Fig.~\ref{Fig2}C that the number of rescue and catastrophe 
events remain invariant along the axon. Their average values 
are smaller than those of the margin of fibroblast cells but 
greater than the fibroblast cell interior. In the next section 
we analytically verify how these differences lead to piecewise 
MT segments of various lengths in axons versus a persistent 
MT growth in bulk and a stable cytoskeleton in fibroblast cells. 

\begin{figure}[t]
\centering
\includegraphics[width=0.8\textwidth]{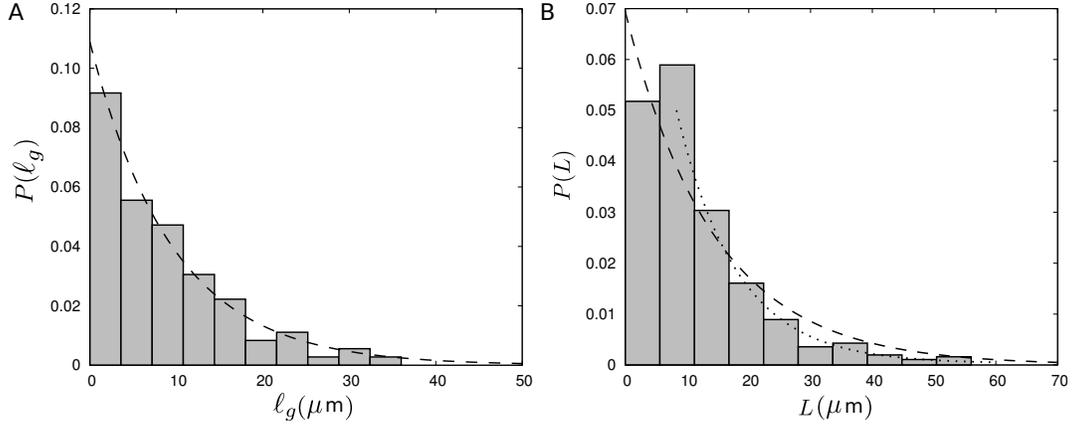}
\caption{(A) The probability distribution of the length $\ell_g$ 
of polymerization episodes. The dashed line is an exponential fit 
$P(\ell_g){\sim}\exp[-\alpha\;\ell_g]$ with $\alpha{=}0.11{\pm}0.01 
\, \mu\text{m}^{-1}$. (B) The estimated probability distribution 
of the MT length $L$. The dashed line corresponds to the analytical 
prediction via Eq.~(\ref{Eq:P-steady}) and the dotted line is an 
exponential fit to the tail.}
\label{Fig3}
\end{figure}

By tracing individual MT trajectories we observe that the tips do 
not experience significant changes in their growth speed until 
the catastrophe occurs (see Fig.~\ref{Fig2}D). Thus, the behavior 
is different from the MT dynamics near the cell margin of 
fibroblast cells. The quick returns to the growth phase due 
to high rescue rate do not exist here, and the shrinkage 
periods are more persistent. The histogram of the growth 
velocity, shown in Fig.~\ref{Fig2}E, is qualitatively similar 
to that of the interior of fibroblast cells. The average growth 
velocity is nearly the same in all samples of axon, yielding an 
overall mean value of $\text{v}_g{=}0.24{\pm}0.05 \, \mu\text{m} 
{/}\text{s}$. The probability distribution $P(\ell_g)$ of the length  
of growth episodes has a mean value of $\langle \ell_g \rangle {\simeq} 
8.7\,\mu\text{m}$ with a decaying tail which can be roughly 
fitted to an exponential curve, as shown in Fig.~\ref{Fig3}A. 

The total length $L$ of MT and its probability distribution $P(L)$ are 
the quantities of interest which can not be directly deduced from our 
experimental results, as the labeling only visualizes the plus ends in 
the growth state and the positions of the minus ends are unknown. 
However, we can indirectly extract an approximate value of the average 
MT length $\langle{L}\rangle$ and the shape of the length distribution, 
by estimating the capacity of the axon tube from the analysis of MT 
trajectories. Denoting the mean number of MTs accommodated in the 
cross-section of the axon by $q$, we start from an arbitrary imaginary 
cross-section and measure the distance $x$ along the axon to reach the 
$q$-th plus-end tip in a given image frame. The average value of the 
fluctuating quantity $x$ corresponds to the half of the typical MT 
length. Thus, by moving the reference cross-section along the axon, 
correcting the data for the variations of the axon thickness, and 
repeating the procedure for all of the image frames, we obtain the 
average MT length $\langle L \rangle{\simeq}11.7\,\mu\text{m}$. 
Moreover, we can construct the probability distribution $P(x)$, 
which is a cumulative distribution function from which $P(L)$ can 
be deduced. The resulting probability distribution $P(L)$ shown in 
Fig.~\ref{Fig3}B develops a small peak at short lengths and has a 
fast decaying tail, indicating that MTs with a length much longer 
that the average value are highly improbable.

\begin{table}[b]
\centering
\caption{MT dynamics parameters.}
\begin{tabular}{cccc}
\addlinespace
\toprule
{\bf } & {Catastrophe frequency} & {Rescue frequency} & {Growth speed} \\
{\bf } & {$f_c\,$ ($\text{s}^{-1}$)} & {$f_r\,$ ($\text{s}^{-1}$)} & 
{$\text{v}_g\,$ ($\mu\text{m} {/}\text{s}$)} \\
\midrule
fibroblast cell &   &  &  \\
interior & $0.004\pm0.002$  & $0.018\pm0.004$  & $0.28\pm0.05$  \\
\midrule
fibroblast cell &   &  &  \\
margin & $0.049\pm0.018$  & $0.114\pm0.027$  & $0.09\pm0.03$  \\
\midrule
axons of human &   &  &  \\
neuron & $0.024\pm0.008$  & $0.019\pm0.006$  & $0.24\pm0.05$  \\
\bottomrule
\end{tabular}
\label{tab:1}
\end{table}

\section*{Model}
\label{Sec:Model}

The MT length regulation mechanism has been theoretically studied over 
the last two decades \cite{Melbinger12,Johann12,Arita15,Ebbinghaus11b,Gonindan04,
Dogterom93}. More recently, the interplay between polymerization kinetics 
and motor-induced depolymerization has been incorporated into the 
stochastic models for the length regulation of MTs and other active 
biopolymers \cite{Melbinger12,Johann12}. However, to predict the filament 
length via these models requires detailed information such as the 
motor concentration on the filament, which is experimentally inaccessible. 
Instead, a conceptually more simple model, proposed by Dogterom and 
Leibler \cite{Dogterom93}, builds on a few parameters that can be 
more easily measured. In this phenomenological model, the evolution 
of the MT length is described in terms of its growth $\text{v}_g$ and 
shrinkage $\text{v}_s$ velocities and the frequencies of catastrophe 
$f_c$ and rescue $f_r$ events. Here we follow such an approach to obtain 
the average steady-state length and its probability distribution as 
well as the conditions under which the filament length diverges.  

We introduce $p_g(L,t)$ and $p_s(L,t)$ as the probabilities of having 
a filament of length $L$ at time $t$ being in the growth or 
shrinkage phase, respectively. One can describe the evolution of 
these probabilities by the following coupled master equations
\begin{equation}
\frac{\partial p_g(L,t)}{\partial t} = f_r \, p_s(L,t) - f_c \, p_g(L,t) 
- \text{v}_g \, \frac{\partial p_g(L,t)}{\partial L},
\label{Eq:Pg}
\end{equation}
\begin{equation}
\frac{\partial p_s(L,t)}{\partial t} = -f_r \, p_s(L,t) + f_c \, p_g(L,t) 
- \text{v}_s \, \frac{\partial p_s(L,t)}{\partial L}.
\label{Eq:Ps}
\end{equation}
While these equations can be analytically solved in general by considering 
appropriate boundary conditions to obtain the time evolution of MT length, 
the steady-state behavior is of the main interest. Therefore, by setting 
the left hand sides of the above equations to zero, after some calculations 
one obtains the probability distribution of the filament length in the 
steady-state as
\begin{equation}
p(\tilde{L}) = p_g(\tilde{L})+p_s(\tilde{L}) = \frac1N \exp[-\ln(k) \, \tilde{L}],
\label{Eq:P-steady}
\end{equation}
where $\tilde{L}$ denotes the dimensionless length of MT in units of 
tubulin dimer length $\delta{\simeq}0.6\,\text{nm}$ (i.e.\ $\tilde{L}{=}L{/}\delta$), 
$N$ is the normalization factor, and $k=\displaystyle\frac{1+f_c\delta/\text{v}_g} 
{1+f_r\delta/\text{v}_s}$. The average MT length at the steady-state follows
\begin{equation}
\langle L \rangle = \int_0^\infty \!\!\!\! p(L) \; L \; dL  \simeq 
\frac{\text{v}_g \, \text{v}_s}{\text{v}_s 
\, f_c - \text{v}_g f_r}.
\label{Eq:Lavg}
\end{equation}
Thus, the steady-state length distribution is governed by the phenomenological 
parameters: the growth and shrinkage rates and the frequencies of catastrophe 
and rescue events. Among the four parameters of the model, $f_c$, $f_r$, and 
$\text{v}_g$ can be directly extracted from the analysis of the live cell 
images. The catastrophe frequency is defined as the number $N_c$ of growth 
to shrinkage transition events (i.e.\ the number of vanishing MTs between two 
frames) over the integrated time $T_g$ spent by MTs in the growth phase. Let 
us assume for simplicity that the imaging frequency is $1\,\text{frame}{/}
\text{s}$. Then, $T_g$ equals to the number $N_g$ of growing MT tips in the 
first frame, and one obtains
\begin{equation}
f_c = \Big\langle \frac{N_c}{T_g} \Big\rangle = \Big\langle \frac{N_c}{N_g} 
\Big\rangle \;\;\; (\text{s}^{-1}),
\label{Eq:f_c}
\end{equation}
where $\langle\cdot\cdot\cdot\rangle$ denotes averaging over all frames. Both 
$N_c$ and $N_g$ can be obtained from the image analysis, which enables us to 
straightforwardly evaluate $f_c$. One can similarly define the rescue frequency 
as
\begin{equation}
f_r = \Big\langle \frac{N_r}{T_s} \Big\rangle = \Big\langle \frac{N_r}{N_s} 
\Big\rangle \;\;\; (\text{s}^{-1}),
\label{Eq:f_r}
\end{equation}
with $N_r$, $T_s$, and $N_s$ being the number of shrinkage to growth 
transition events (i.e.\ new MTs), the integrated time spent by MTs in 
the shrinkage phase, and the total number of shrinking MT tips in the 
first frame, respectively. However, $N_s$ is not directly accessible 
since EB1 labeling technique only visualizes the dynamics in the growth 
phase. From $N_g{=}\frac{f_r}{f_r{+}f_c}N$ and $N_s{=}\frac{f_c}{f_r{+}
f_c}N$ ($N$ denotes the total number of MTs in a frame), we get $\frac{N_g}{N_s} 
{=}\frac{f_r}{f_c}$. From this relation and Eq.~(\ref{Eq:f_r}), we obtain 
$f_r$ by a two-parameter fit. The results are given in table \ref{tab:1}. 
For the remaining parameter, i.e.\ the shrinkage rate $\text{v}_s$, we 
take $\text{v}_s\sim0.59\pm0.21\,\mu\text{m}{/}\text{s}$ from the 
literature \cite{Komarova02b}, as we do not expect that $\text{v}_s$ 
is considerably affected by the presence of obstacles or in the vicinity 
of the cell margin. From Eq.~(\ref{Eq:Lavg}) one obtains $\langle L \rangle 
\simeq 15.7 \,\mu\text{m}$ in a good agreement with the experimental value 
$\langle L \rangle{\simeq}11.7\,\mu\text{m}$, despite all the approximations 
made in the model as well as the inaccuracies in evaluating the quantities 
of interest in experiments. We also compare the analytical prediction of 
$p(L)$ from Eq.~(\ref{Eq:P-steady}) with the distribution estimated from 
the experimental data. As shown in Fig.~\ref{Fig3}B, there is a remarkable 
agreement, even though the small peak is not captured.

\begin{figure}[t]
\centering
\includegraphics[width=0.6\textwidth]{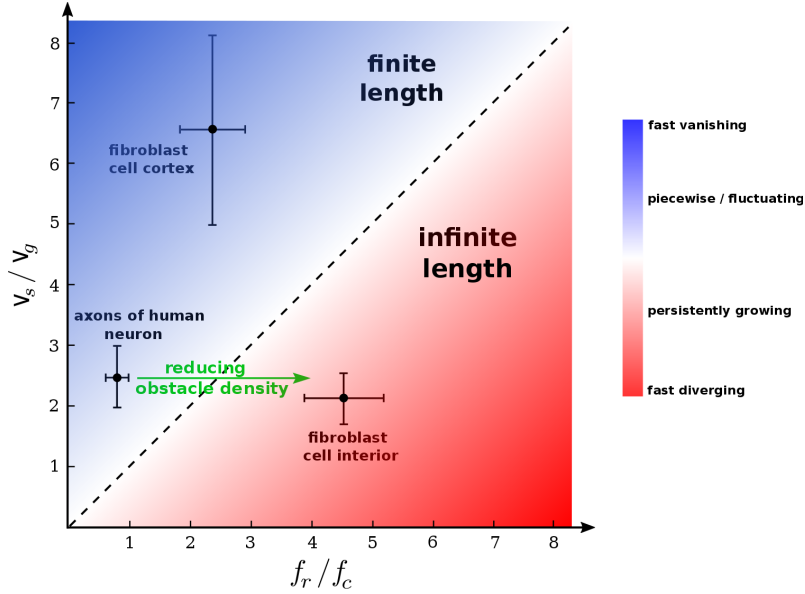}
\caption{Phase diagram of MT length regulation in the $\text{v}_s{/}\text{v}_g 
- f_r{/}f_c$ plane. The dashed line indicates unity and separates the 
steady-state diverging and finite length regimes. The color intensity 
indicates possible scenarios of MT dynamics.}
\label{Fig4}
\end{figure}

\section*{Discussion}
We investigated the altering phases of polymeriztion and depolymerization 
of MTs in different cell types and regions, and showed that the dynamics 
of MTs in fibroblast cells is distinctive from neuronal axons. The behavior 
even differs considerably between the interior and margin of fibroblast 
cells, i.e.\ after reaching from the cell interior all the way to the plasma 
membrane, the growth dynamics of MTs significantly changes. Importantly, 
MT tip fluctuations and collecting more active tips near the membrane 
enables the cell to quickly respond to changes in the environmental 
conditions, adapt its shape, or advance its edge and move. In contrast, 
the tip experiences a persistent growth/shrinkage phase within the cell 
interior, which results in a rather stable filamentous structure in the 
bulk of the cell. The combination of the two types of MT dynamics in 
bulk and margin allows the cell to maintain the connection between 
nucleus and membrane.  

The growth velocity of MT tips in our experiments, even in the margin of 
fibroblast cells, is far greater than those reported for MT growth against 
rigid obstacles \cite{Dogterom97,Dogterom02,Dogterom05} or those with 
coated beads coupled to their ends \cite{Trushko13}, thus, these conditions 
were different and we cannot reasonably consider the force-speed relations 
that were determined in these works for interpretation of our observations. 
The persistent intracellular growth followed by a highly unstable dynamics 
near the cell margin was also reported in other eukaryotic cells, such as 
CHO-K1 cells \cite{Komarova02b}, which was attributed to the promotion of 
rescue rate near the membrane induced e.g.\ by the presence of CLIP-170 
linker proteins \cite{Komarova02b}. These proteins also enable MTs to
distinguish different cortical regions and regulates their catastrophe 
rate accordingly \cite{Brunner00}. Moreover, it has been shown that 
barrier-attached dynein can inhibit MT growth and trigger microtubule 
catastrophes \cite{Laan12}. The lack of GTP in the vicinity of the 
membrane, and also the spatial variations of the distribution of 
mitochondria can be other influential factors to the dynamics of MTs 
at different cell regions. Understanding the underlying mechanisms 
of MT dynamics regulation is crucial and requires further detailed 
studies, which is the subject of our ongoing research.  

From the phenomenological model of MT length regulation one obtains a 
criterion for the transition from finite length to diverging MTs. It 
can be deduced e.g.\ from Eq.~(\ref{Eq:Lavg}) that the MT length 
diverges if $\frac{f_r}{f_c} {>} \frac{\nu_s}{\nu_g}$. Therefore, we summarize 
the MT length regulation in a phase diagram in the $\nu_s{/}\nu_g - f_r{/}
f_c$ plane in Fig.~\ref{Fig4}. While the parameter values corresponding 
to the interior of fibroblast cells are located in the diverging regime 
of the phase diagram, the model successfully predicts a finite steady-state 
length for both the fibroblast cortex and the axons of neurons. A 
comparison between axon and fibroblast cell interior reveals that 
the parameter values are very similar, except for the catastrophe frequency 
$f_c$ which shows a sixfold increase in axons. A plausible scenario is 
that MT parameter values in axons are designed for an infinite growth 
along the tube, similar to the bulk of eukaryotic cells. However, 
the frequency of transition from growth to shrinkage phase increases when 
growing against large obstacles in laterally-limited crowded tubes of 
axons, which causes a piecewise MT segment structure. Thus, it is expected that 
$f_c$ decreases and MTs grow unlimitedly with reducing the obstacle 
density in the axon tube. Further investigations of the dynamics and 
spatial organization of MTs in axons remain for future work, which is 
a crucial step towards a better understanding of the underlying mechanisms 
of bidirectional transport driven by cytoskeletal motors \cite{Appert-Rolland15} 
and the impact of neurodegenerative disorders on it. More generally, 
the efficiency of the intracellular transport substantially depends on 
the structure of the cytoskeleton \cite{Shaebani14,Sadjadi15}, which 
underlines the need of detailed studies in other cell types to understand 
the structural dynamics of cytoskeletal biopolymers and their spatial 
variations with respect to the cell boundaries. As a final remark, 
we observed no aging effects in MT dynamics in the biological systems 
under consideration within the temporal resolution of our experiments, 
which allowed us to adopt a simple set of master equations (\ref{Eq:Pg}) 
and (\ref{Eq:Ps}) to describe the MT dynamics. In general, however, 
the MT dynamics can be age and length dependent \cite{Stepanova10,Gupta06}. 
In such cases, the formalism can be generalized by explicitly including 
the time and length dependence of the phenomenological parameters. It is 
also possible to analytically handle the spatial and temporal variations 
of the free tubulin concentration which influences the MT dynamics 
\cite{Dogterom93,Janulevicius06}.

\section*{Acknowledgments}
We are grateful to Prof.\ Gerald Thiel for providing SH-SY5Y cells. 
pGFP-EB1 was a gift from Lynne Cassimeris (Addgene plasmid \# 17234). 
We acknowledge useful discussions with Drs Ksenia Astanina, Konstantin 
Lepikhov and Chikashi Arita, and technical support by Javad Najafi. 
This work was funded by the Deutsche Forschungsgemeinschaft (DFG) 
through Collaborative Research Center SFB 1027 (Projects A7 and C1).

\section*{Author contributions statement}
All authors designed and conducted the research, AP performed the experiments, 
MRS and LS developed the theoretical framework, MRS analyzed data and wrote 
the paper.

\section*{Additional information}
{\bf Competing financial interests:} The authors declare no competing 
financial interests.

\end{document}